\documentclass[superscriptaddress,aps,amsfonts,notitlepage,twocolumn,nofootinbib]{revtex4-2}
\usepackage{xcolor,graphicx}

\usepackage{breqn,amsmath,amssymb}
\usepackage{footmisc}

\usepackage{enumerate}

\usepackage{booktabs}
\usepackage{makecell}

\usepackage{comment}

\usepackage{subcaption}

\usepackage[T1]{fontenc}
\usepackage{natbib,hyperref}

\usepackage{mathtools}

\usepackage{centernot}

\makeatletter
\let\cat@comma@active\@empty
\makeatother

\usepackage{IEEEtrantools}

\begin{document}

\title{Luminal Scalar-Tensor theories for a not so dark Dark Energy}

\author{S. Mironov}
\email{sa.mironov\_1@physics.msu.ru}
\affiliation{Institute for Nuclear Research of the Russian Academy of Sciences, 
60th October Anniversary Prospect, 7a, 117312 Moscow, Russia}
\affiliation{Institute for Theoretical and Mathematical Physics,
MSU, 119991 Moscow, Russia}
\affiliation{NRC, "Kurchatov Institute", 123182, Moscow, Russia}

\author{A. Shtennikova}
\email{shtennikova@inr.ru}
\affiliation{Institute for Nuclear Research of the Russian Academy of Sciences, 
60th October Anniversary Prospect, 7a, 117312 Moscow, Russia}
\affiliation{Institute for Theoretical and Mathematical Physics,
MSU, 119991 Moscow, Russia}

\author{M. Valencia-Villegas}
\email{mvalenciavillegas@itmp.msu.ru}
\affiliation{Institute for Theoretical and Mathematical Physics,
MSU, 119991 Moscow, Russia}
\affiliation{Institute for Nuclear Research of the Russian Academy of Sciences, 
60th October Anniversary Prospect, 7a, 117312 Moscow, Russia}

\begin{abstract}
In general the speed of Gravitational Waves (GWs) in Scalar-Tensor modifications of Einstein's gravity is different from the speed of Light. Nevertheless, it has been measured that their speeds are nearly the same. For the most general Scalar-Tensor theories classified to date that do propagate a graviton --- DHOST, {\it including Horndeski and Beyond Horndeski (BH) theories} --- we show that, remarkably, up to 5 self-consistent couplings of  the scalar of Dark Energy (DE) to the Photon are enough to make their GWs luminal in a wide set of cases. 
 We find at least one Luminal BH theory for which the GW decay into DE is suppressed in any cosmological background.
\end{abstract}

\maketitle

With the new era of multi-messenger astronomy, the initial impression was that a large class of Scalar-Tensor modifications of Einstein's gravity is ruled out by the strict coincidence between the speed of Light ($c$) and Gravity ($c_g$). In particular, the almost simultaneous detection of the Gravitational Wave (GW) signal from the event GW170817 \cite{LIGOScientific:2017vwq} and the gamma ray Burst GRB170817A \cite{Goldstein:2017mmi,Savchenko:2017ffs} placed the strong constraint, 
\begin{equation}
    \left\vert \frac{c_g}{c} -1\right\vert \leq 5\times 10^{-16}\,. \label{eqn speedtest}
\end{equation}
Nevertheless, it is clear that this constraint essentially indicates a relation between Gravity and Light. It, nevertheless, does not directly rule out modified gravity theories that could be relevant on cosmological scales, without assuming in first place something about light also on those scales. The standard approach is to take the following assumption:
\begin{enumerate}[(b)]
    \item{The Photon of Maxwell Electrodynamics (EM) remains minimally coupled even at the scales where General Relativity (GR) may need modification. Namely, $c=1$ even at the scales where the scalar of Dark Energy dominates the expansion of the universe,}
\end{enumerate}
thus, we see $c_g=c=1$. However, {\it gravity couples universally to all matter}, and in principle, one could also explore an alternative assumption to (b), where the {\it scalar modification of gravity} at cosmological scales shares this {\it universal coupling} property, {\it e.g.}:
\begin{enumerate}[(a)]
        \item{The scalar of Dark Energy (DE) couples to both, the Graviton and the Photon in a specific way, {\it such that we see the luminality of GWs}}\label{item notsodark}
\end{enumerate}
 \begin{equation}
        \frac{c_g(t)}{c(t)}=1\,. \label{eqn unitratio}
    \end{equation}
The assumption (b) conveniently fits EM at {\it all} scales, mainly to be consistent with laboratory experiments. However, it also quickly forbids additional input to more objectively constrain modified gravity theories.

The assumption (a) --- which we take in this letter, and that clearly contains (b) as a  particular case --- opens a new set of observational possibilities: If DE is not really dark and also couples to the Photon, new types of laboratory and astrophysical tests are required \cite{Brax:2014vva}. 

Furthermore, (a) re-opens the path to non-minimally coupled theories for DE that were previously thought to be ruled out  \cite{Mironov:2024idn,Bettoni:2016mij,Ezquiaga:2017ekz,Sakstein:2017xjx,Baker:2017hug,Creminelli:2017sry, Langlois:2017dyl}. Interestingly, non-minimal couplings may be relevant in the wake of the recent DESI BAO data, favoring dynamical DE \cite{DESI:2024mwx}. To that end a Horndeski  theory with non-minimal couplings of the scalar to gravity would be necessary to safely cross the phantom divide \cite{Ye:2024ywg,Wolf:2024eph} (See however the discussions in {\it e.g.} \cite{Colgain:2024mtg,Colgain:2024xqj,Shlivko:2024llw,Cortes:2024lgw,Patel:2024odo}). Although these possibilities are not  conclusive \cite{Wolf:2024stt}, the theories shown in this letter --- a broad generalization of the theories  used in \cite{Ye:2024ywg,Wolf:2024eph}--- open new opportunities for the cosmologist.

We consider the most general  Degenerate Higher-Order Scalar-Tensor modifications of gravity (DHOST) that have been classified to date, which are by construction free of Ostrogradsky ghosts  \cite{Langlois:2015cwa,BenAchour:2016fzp,Langlois:2017mxy,Langlois:2018dxi, Mironov:2024pjt}. We deduce the DE--Photon couplings that are necessary for the observed luminality of GWs in these DHOST (\ref{eqn unitratio}) --- with {\it Horndeski and Beyond Horndeski} as particular cases. We find that only two types of DE--Photon couplings are necessary. One of them cannot be removed by a conformal/ disformal transformation of the metric. It is involved in a new Luminal Beyond Horndeski (BH) theory that we show below, for which the GW decay to DE  is suppressed. Altogether passing the strong constraints  on both the Luminality \cite{Bettoni:2016mij,Ezquiaga:2017ekz,Sakstein:2017xjx,Baker:2017hug,Creminelli:2017sry, Langlois:2017dyl} and non-decay of GW \cite{Creminelli:2018xsv}.\\ 

{\it The model:} In the usual parameterization, we consider 19 potentials depending on a scalar field $\pi$. They generalize the Einstein-Hilbert action in four dimensions (4D) with minimal and non-minimal couplings of $\pi$ to gravity. Let us denote the 19 scalar potentials as $a_i,\, b_j,\, f_k,\, G_k$ with $i=1\, \dots 5,\, j=1\, \dots 10$ and $k=2,3$. In principle, we allow all of these potentials to be functions of a scalar field $\pi$ and  $X=\pi_\mu\, \pi^\mu$, where $\pi_\mu=\nabla_\mu\pi$. However, some of these potentials are not free. There are relations among them in order to not propagate the Ostrogradsky ghost. These relations, known as degeneracy conditions, separate the theory space of DHOST into distinct classes. Thus, in all the theories we consider there are always less than 19 free scalar potentials of $\pi$ and $X$, with the specific number of free functions depending on the class. For instance in Horndeski theory there are up to $4$ free functions \cite{horndeski1974second,Deffayet:2011gz}. A complete classification with the number of free functions, and  properties is given in \cite{BenAchour:2016fzp,Langlois:2017mxy,Langlois:2018dxi}. Below we only give the degeneracy conditions for the most physically relevant cases.

The Lagrangian is written as
\begin{equation}
    \resizebox{1\hsize}{!}{ $\mathcal{L}_{\text{DHOST}_{\pi}}=f_2\,R+f_3\,G_{\mu\nu}\, \pi^{\mu\nu}+\mathcal{L}_{\text{Linear}}+\mathcal{L}_{\text{Quad}}+\mathcal{L}_{\text{Cubic}}\,,$} \label{eqn dhostL}
\end{equation}
where $R$ is the Ricci scalar, $G_{\mu\nu}$ is the Einstein tensor and $\nabla$ is the covariant derivative computed with the ambient metric of the $D$-dimensional manifold (of signature $-,+,+,+,\,\dots$), and $\mu=0,1,\,\dots\, D-1$. The main results in this letter will be in the usual $D=4$, however, as we explain latter on, they are most easily derived starting from $D=5$, as we will explicitly state when needed. 

The  last three terms in (\ref{eqn dhostL}) contain diffeomorphism invariant combinations of respectively, (up-to) linear, quadratic and cubic in $\nabla^2\pi$ terms. Explicitly,
\begin{equation}
    \mathcal{L}_{\text{Linear}}=G_2+G_3\,\Box \pi 
\end{equation}
 \begin{equation}
        \mathcal{L}_{\text{Quad}}=\sum_{i=1}^5\,a_i(\pi,X) \,L_i^{(2)}\,, \label{eqn Lquad}
    \end{equation}
    where $L_i^{(2)}$ are of order  $(\nabla^2\pi)^2$,
\begin{center}
 \begin{tabular}{ccc}
      $L_1^{(2)}=(\pi_{\mu\nu})^2\,,$ & $L_2^{(2)}=(\Box\pi)^2\,,$ & $L_3^{(2)}=\Box\pi\,(\pi_{\mu\nu}\pi^\mu \pi^\nu)\,,$
    \end{tabular}
    \begin{equation}
       L_4^{(2)}=(\pi_{\mu\rho}\pi^{\mu})^2\,,\,\,L_5^{(2)}=(\pi_{\mu\nu}\pi^{\mu}\pi^{\nu})^2\,,
   \end{equation}
\end{center}
and 
\begin{equation}
    \mathcal{L}_{\text{Cubic}}=\sum_{j=1}^{10}\,b_j(\pi,X) \,L_j^{(3)}\label{eqn Lcubic}
\end{equation}
where $L_j^{(3)}$ are of order  $(\nabla^2\pi)^3$,
\begin{center}
 \begin{tabular}{ccc}
      $L_1^{(3)}=(\Box\pi)^3\,,$ & $L_2^{(3)}=\Box\pi(\pi_{\mu\nu})^2\,,$ & $L_3^{(3)}=(\pi_{\mu\nu})^3\,,$
    \end{tabular}\bigskip
   
    \begin{tabular}{cc}
      $L_4^{(3)}=(\Box \pi)^2 (\pi_{\mu\nu}\pi^{\mu}\pi^{\nu})\,,$ & $L_5^{(3)}=\Box\pi\,(\pi_{\mu\nu}\pi^{\mu})^2\,,$ \\
      &\\
      $L_6^{(3)}=(\pi_{\rho\sigma})^2\,(\pi_{\mu\nu}\pi^{\mu}\pi^\nu)\,,$ & $L_7^{(3)}=\pi^{\mu\nu}\pi_{\nu\rho}\pi^{\rho\sigma}\pi_{\mu}\pi_\sigma\,,$ \\
      &\\
      $L_8^{(3)}=(\pi^{\mu\nu}\pi_\mu)^2(\pi^{\rho\sigma}\pi_{\rho}\pi_{\sigma})\,,$ & $L_9^{(3)}=\Box\pi (\pi^{\rho\sigma}\pi_{\rho}\pi_{\sigma})^2\,,$
    \end{tabular}
    \begin{equation}
    L_{10}^{(3)}=(\pi^{\rho\sigma}\pi_{\rho}\pi_{\sigma})^3
    \end{equation}
\end{center}

It was thought that a large set of DHOST theories -- including Horndeski and Beyond Horndeski \cite{Gleyzes:2014dya, Gleyzes:2014qga} -- are constrained to some extent in order to satisfy (\ref{eqn unitratio}). As we noted, this belief assumed (b). In this letter, however, we work on the hypothesis (a), and thus, to $\mathcal{L}_{\text{DHOST}_\pi}$ we must add the precise Scalar of DE--Photon couplings such that we see the luminality of GWs (\ref{eqn unitratio}). It was initially shown in \cite{Mironov:2024idn} that a simple way to obtain them in 4D is to start from a 5 Dimensional (D) setup: thus, consider the action of DHOST only for a brief moment in 5D,
\begin{equation}
    \int \sqrt{-{}^{(5)}g }\, d^5\text{x}\, \mathcal{L}_{\text{DHOST}_\pi}\,.\label{eqn 5Daction}
\end{equation}
Writing the 5D metric ${}^{(5)}g$ as
\begin{equation}
{}^{(5)}g_{B\,C}=\left(
\begin{array}{cc}
g_{\mu\,\nu}+\,A_\mu\,A_{\nu} & \,A_\mu  \\  
\,A_\nu & 1
\end{array}\right)\,, \label{eqn kaluzametric}
\end{equation}
where the latin indices are $B=0,\,\dots \, 4$ and greek $\mu=0,\,\dots \, 3$. Seen simply as a tool for our purpose in 4D, we compactify the 5th dimension with Kaluza's cylinder condition \cite{Kaluza:1921tu}, where we have assumed right away in Eqn. (\ref{eqn kaluzametric}) a constant Dilaton, and such that the 4D fields $g$ and $A_\mu$ do not depend on the 5-th dimension. We further rescale the 4D fields to re-absorb the $\int d\text{x}^4$, and thus we rewrite (\ref{eqn 5Daction}) in terms of 4D fields only. 

All in all, after compactification, the theory (\ref{eqn 5Daction}) takes the form of the usual DHOST plus a Scalar--Photon sector in 4D   (\ref{eqn actionF}). As usual  the $U(1)$ gauge invariance in the vector sector is inherited from diffeomorphisms in 5D.

It is clear that because the 4-vector $A_\mu$ and the 4D metric are just but components of the same metric in 5D, their speed in 4D is generally bound to be {\it the same}. The caveat is that we have broken isotropy in 5D by compactifying one spatial dimension and ignoring\footref{foot dilaton} the dynamics of a Dilaton. Thus, there are special cases with unequal speeds which we single out below.\\

{\normalsize {\bf DHOST with Dark Energy--Photon couplings:}} From now on in 4D, the complete DHOST action with DE--Photon couplings  reads,
\begin{eqnarray}
     \int \sqrt{-g }\, d^4\text{x} \left(\mathcal{L}_{\text{DHOST}_{\pi}}+\mathcal{L}_{\text{DHOST}_{A}}\right)\,,\label{eqn actionF}
\end{eqnarray}
with $\mathcal{L}_{\text{DHOST}_{\pi}}$ given in (\ref{eqn dhostL}). The DE--Photon sector is,
\begin{eqnarray}
    \mathcal{L}_{\text{DHOST}_{A}}&=&
    \frac{f_3}{8}\left(4F_{\mu\nu}\nabla_\rho F^{\nu\rho} \pi^\mu+F^2\Box\pi-4F_\mu{}^\nu F^{\mu\rho}\pi_{\nu\rho} \right)\nonumber\\
    &-&\frac{f_2}{4}\, F^2+l_{\text{Quad}_A}+l_{\text{Cubic}_A}\,, \label{eqn dhostLA}
\end{eqnarray}
where $F_{\mu\nu}=\partial_\mu A_\nu-\partial_\nu A_\mu$, and with obvious notation parallel to (\ref{eqn Lquad}) and (\ref{eqn Lcubic}),
\begin{eqnarray}
 l_{\text{Quad}_A}&=&a_1(\pi,X) l_1^{(2)}\label{eqn l12}\\
    l_{\text{Cubic}_A}&=&\sum_{j=\{2,3,6\}} b_j(\pi,X) l_j^{(3)}\,,
\end{eqnarray}
where $l_1^{(2)}=l_j^{(3)}=\frac{1}{2}  (F_{\mu\nu}\pi^\mu)^2 $ for $j=2,6$, and 
\begin{equation}
    l_3^{(3)}=\frac{3}{4}  F_{\mu\nu}F_{\rho\sigma}\pi^{\mu\rho}\pi^\nu\pi^\sigma\,.
\end{equation}
One identifies in principle three types of DE--Photon couplings $F^2\, \nabla^2\pi,\, F^2\,(\nabla\pi)^2$ and $F^2\,\nabla^2\pi\,(\nabla\pi)^2$. However, the latter --- proportional to $b_3$ --- will be removed below by the Luminality condition (\ref{eqn unitratio}). It is essential to note that the DE--Photon couplings $ f_3\,  F^2 \nabla^2 \pi$ {\it cannot be removed by a conformal/ disformal transformation} of the metric that depends on up to first derivatives of $\pi$. Namely, one cannot obtain the $f_3$ DE--Photon couplings by such metric redefinition in the Maxwell term $-1/4 \, F^2$.

In short, for each of the 6  contributions to $\mathcal{L}_{\text{DHOST}_{\pi}}$ labeled by the scalar potentials $f_2,\, f_3,\,a_1,\,b_2,\,b_3,\,b_6$ there is a corresponding DE--Photon sector in Eqn. (\ref{eqn dhostLA}).\\  

{\bf  Luminal DHOST around the corner:}  As declared, not all Lagrangians in Eqn. (\ref{eqn actionF}) propagate gravitational and electromagnetic waves at the same speed. The approach below is to compute the Graviton and Photon speeds on a cosmological  background for the action (\ref{eqn actionF}) and find the Lagrangians that can satisfy Eqn. (\ref{eqn unitratio}). Note that previous cases suggest that the results below could also hold on (at least)  spherically symmetric  backgrounds \cite{Mironov:2024yqa,MironovNew}.

The scalar mode of DHOST is {\it not modified} by the new terms $\mathcal{L}_{\text{DHOST}_{A}}$ on the cosmological background. Thus we do not discuss any further the scalar sector in this letter. Furthermore, we will assume the DHOST classes that actually propagate a graviton  \cite{Langlois:2015cwa,BenAchour:2016fzp,Langlois:2017mxy,Langlois:2018dxi, Mironov:2024pjt}.

We consider first order perturbations on a spatially flat FLRW background. With the perturbed metric $\text{d}s^2=(\eta_{\mu\nu}+\delta g_{\mu\nu})\text{d}x^\mu \text{d}x^\nu$ where $\eta_{\mu\nu}=-\text{d}t^2+a(t)^2\,\delta_{ij}\text{d}x^i \text{d}x^j$, we write only the symmetric, traceless and transverse tensor perturbation $h_{ij}$ and the two transverse vector perturbations $S_i,\, F_i$ as,
\begin{equation}
\delta g=\left(2\,S_i \textrm{d}t \, \textrm{d}x^i+\left(\partial_i F_j+\partial_j F_i+2\,h_{ij}\right) \textrm{d}x^i \, \textrm{d}x^j \right)\,,
\end{equation}
where we denote spatial indices with lowercase latin indices, $i=1,2,3$. The perturbed DHOST scalar $\pi(x^\mu)$ is written as $\pi(t)+\chi(t,\vec{x})$ in the linearized expressions, within which $\pi(t)$ is the background scalar field. Finally, on the cosmological medium the photon amounts to the transverse perturbation $A_i(t,\vec{x})$, with vanishing background due to isotropy.

The quadratic action for the graviton reads,
\begin{equation}
  \mathcal{S}_{Tensor}=\frac{1}{2}\int  \text{d}t\,\text{d}^3x\,a^3 \,\left(\mathcal{G}_\tau\,\dot{h}_{ij}^2-\frac{\mathcal{F}_\tau}{a^2}\,(\partial_k h_{ij})^2\right)\,,
\end{equation}
with $f_{3,X}=\frac{\partial f_3}{\partial X}$ and so on,
\begin{eqnarray}
    \mathcal{G}_\tau&=& 2f_2+2\ddot{\pi}X f_{3,X}-Xf_{3,\pi} -2Xa_1\\
    &+&2X(3\dot{\pi}H+\ddot{\pi})b_2+6\dot{\pi}X H b_3+2\ddot{\pi}X^2 b_6\,,\nonumber\\
     \mathcal{F}_\tau&=&2f_2-2\ddot{\pi}X f_{3,X}+X f_{3,\pi}\,,
\end{eqnarray}
while the action for the Photon is written as,
\begin{equation}
  \mathcal{S}_{Vector}=\frac{1}{4}\int  \text{d}t\,\text{d}^3x\,a \,\left(\mathcal{G}_A\,\dot{A}_{i}^2-\frac{\mathcal{F}_A}{a^2}\,(\partial_k A_{i})^2\right)\,,
\end{equation}
where,
\begin{eqnarray}
    \mathcal{G}_A&=& \mathcal{G}_\tau-3 \dot{\pi} X H b_3 \label{eqn GA} \\
     \mathcal{F}_A&=&\mathcal{F}_\tau\,.\label{eqn FAFT}
\end{eqnarray}
As expected, by construction, the coefficients in the  quadratic actions are similar, {\it e.g.} as in (\ref{eqn FAFT}). Now, with their speeds squared, respectively, $c_g^2=\frac{\mathcal{F}_\tau}{\mathcal{G}_\tau}$ and $c^2=\frac{\mathcal{F}_A}{\mathcal{G}_A}$, we find their ratio, 
\begin{equation}
    \frac{c_g^2}{c^2}=1-3\frac{\dot{\pi}X H\,  b_3}{\mathcal{G}_\tau}\, .
\end{equation}
Thus, in principle, the DHOST theories with
\begin{equation}
    b_3=0\,,\label{eqn luminalityCondition}
\end{equation}
would preserve the unit ratio of speeds (\ref{eqn unitratio}). However, let us recall that depending on the degenerate class of DHOST being considered, the scalar potential $b_3$ {\it may not be a free function}  but it may be fixed  by the also crucial degeneracy conditions\footnote{If we had considered the Dilaton, with background $\Phi(t)$, then $\frac{c_g^2}{c^2}=1-3\frac{\dot{\pi}X (H-\frac{\dot{\Phi}}{\Phi})\Phi\,  b_3}{\mathcal{G}_\tau}$. Restoring isotropy $\dot\Phi=\dot{a}$, we would see that (\ref{eqn unitratio}) always holds. Although this choice is {\it unphysical}, this is at the very least a cross-check of our results. See \cite{Mironov:2024umy} Section VC for a discussion. \label{foot dilaton}}.\\

{\bf Degenerate and Luminal DHOST:} thus, to apply the luminality condition (\ref{eqn luminalityCondition}) in DHOST, one is left with the task of establishing whether it is consistent with the degeneracy condition of the class. From the comprehensive classification in \cite{Langlois:2017mxy} Table 1 and \cite{BenAchour:2016fzp} it is clear that there are many\footnote{A counterexample, where the luminality is not compatible with the degeneracy condition, is the {\it full} mixed quadratic plus cubic BH \cite{Gleyzes:2014dya, Gleyzes:2014qga}. Let us see: the degeneracy condition is (\ref{eqn degCondition}). While  $b_3=0$ implies $F_5=-\frac{G_{5,X}}{3 X}$. {\it Assuming} $F_5,\,G_{5,X}\neq0$ one finds from (\ref{eqn degCondition}) a relation $F_4(G_4,\,G_{4,X},\,G_{5,\pi})$  that sets $\mathcal{G}_\tau=\mathcal{G}_A=0$, which is a singular case with no Graviton and no Photon. 

However, note that the branch $F_5=G_{5,X}=0$ escapes the problem, because (\ref{eqn degCondition}) is automatically satisfied with a totally free $F_4$. See the discussion below in the case (ii).

Another counterexample is only cubic, full DHOST ${}^3 \text{N-I}$, which contains cubic Horndeski and BH: as noted in \cite{BenAchour:2016fzp} in this class $b_3=2 b_1$, with $b_1$ free, up to the condition $b_1\neq 0$. Thus in this class (\ref{eqn luminalityCondition}) cannot be met. If we nevertheless take $b_1=0$, then we would be forced in another degenerate class, DHOST ${}^3 \text{N-II}$ \cite{BenAchour:2016fzp}, which however has no graviton \cite{Langlois:2017mxy}. \label{foot ce}} Scalar-Tensor theories with a graviton that can be made Luminal with Eqn. (\ref{eqn luminalityCondition}). We will focus, however, on the phenomenologically most relevant classes. The simplest successful case is:

\begin{enumerate}[(i)]
    \item Every {\bf quadratic DHOST with the corresponding DE--Photon couplings}, $\mathcal{L}_{\text{DHOST}_{A}}=-\frac{f_2}{4}\,F^2+\frac{a_1}{2} (F_{\mu\nu}\pi^\mu)^2$,  and with a graviton, satisfies  Eqn. (\ref{eqn unitratio}). Namely, the action (\ref{eqn actionF}) with $b_i=0$ and $f_3=0$ has luminal GWs. The degeneracy conditions on some of the functions $a_i$, for multiple classes of theories, are given for instance in \cite{BenAchour:2016fzp} App. C.
\end{enumerate}

In particular, (i) includes {\it quadratic Horndeski and Beyond Horndeski theory} ($\text{BH}_{4}$) as special cases. The latter is written with the action (\ref{eqn actionF}) and with the following degeneracy relations \cite{Langlois:2015cwa,Langlois:2018dxi},
$$f_2=G_4,\, a_1=-a_2=2G_{4,X}+X F_4,\, $$
\vspace{-0.8cm}
\begin{equation}
    a_3=-a_4=2 F_4,\,a_5=0\,.\label{eqn quadBH}
\end{equation}
From (\ref{eqn dhostLA}) and (\ref{eqn quadBH}) the DE--Photon couplings that make luminal the quadratic BH theory are $\mathcal{L}_{BH_{4A}}$ in Eqn. (\ref{eqn lagrangianBH4H5}), from which we recover $\mathcal{L}_{4A}$ given in \cite{Mironov:2024idn} in the particular case $F_4=0$. The theory (i) also includes as a luminal class, for instance, the DHOST ${}^2 \text{N-III/ IIa}$, which may still be phenomenologically relevant \cite{Langlois:2017mxy}, yet disconnected from the Horndeski class.\\

Another successfully Luminal case is in the mixed quadratic plus cubic DHOST class:
\begin{enumerate}[(ii)]
    \item The {\bf Quadratic Beyond Horndeski (BH) plus Cubic Horndeski theory, with $f_3=G_5(\pi)$ and with the corresponding DE-Photon couplings} propagates Luminal GWs. Namely, the action (\ref{eqn actionF}) with the relations (\ref{eqn quadBH}), and with $b_j=0$ with $j=1,\dots 10$. Explicitly,
\begin{equation}
\int \text{d}^4x \big(\,\mathcal{L}_{BH_{4\pi}}+\mathcal{L}_{BH_{4A}}+\mathcal{L}_{H_{5\pi}}+\mathcal{L}_{H_{5A}}\big)\,,\label{eqn lagrangianBH4H5}
\end{equation}
with
\begin{eqnarray}
    &&\mathcal{L}_{BH_{4\pi}}=G_2+G_3 \Box\pi+G_4 R-2 G_{4,X}((\Box\pi)^2-\pi_{\mu\nu}^2)\nonumber\\
    &&-F_4\Big(X (\Box \pi)^2-X \pi_{\mu\nu}^2+2 (\pi_{\mu\nu}\pi^\mu)^2-2 \Box \pi \pi_{\mu\nu} \pi^\mu \pi^\nu\Big)\nonumber\\
    &&\nonumber\\
   && \mathcal{L}_{\text{BH}_{4A}}=-\frac{G_4}{4}\,F^2+\frac{2G_{4,X}+X F_4}{2} (F_{\mu\nu}\pi^\mu)^2\,,\nonumber\label{eqn BH4A}\\
    &&\nonumber\\
   && \mathcal{L}_{H_{5\pi}}=G_5\, G^{\mu\nu}\pi_{\mu\nu}\\
   &&\nonumber\\
  && \mathcal{L}_{H_{5A}}= \frac{G_5}{8}\left(4F_{\mu\nu}\nabla_\rho F^{\nu\rho}\pi^\mu+F^2\Box\pi-4F_\mu{}^\nu F^{\mu\rho}\pi_{\nu\rho} \right)\,. \nonumber
\end{eqnarray}
$G_5$ is a function of $\pi$ only, and we have taken $F_5=0$ (in the standard notation of BH \cite{Langlois:2018dxi}). 
\end{enumerate}

The theory (ii) generalizes the Luminal Horndeski theory with $G_4(\pi,X),\,G_5(\pi)$ shown in \cite{Mironov:2024idn} to also include $F_4(\pi,X)$. This is {\it essential}: namely, with this new theory it becomes possible  to suppress the GWs decay to the scalar of DE, by fixing the newly free potential $F_4$. Let us see how: First note that $b_3=\frac{2}{3}(G_{5,X}+3X F_5)=0$ is satisfied. Then Eqn. (\ref{eqn unitratio}) follows; that is, in the theory (ii) the GWs are automatically Luminal {\it without fixing any of the scalar potentials}. Secondly, this theory is free of Ostrogradsky ghosts: the degeneracy condition\footnote{We take sign convention for $F_4,\, F_5$ from \cite{Langlois:2015cwa,Langlois:2018dxi}. Note however, the opposite sign for $F_4$ taken in \cite{Creminelli:2017sry,Creminelli:2018xsv,Babichev:2024kfo}}  in mixed quadratic plus cubic BH
\begin{eqnarray}
    F_4 G_{5,X} X=-3F_5 \Big(G_4-2 X G_{4,X}-\frac{X}{2}G_{5,\pi}\Big),\,\label{eqn degCondition}
\end{eqnarray}
is also automatically satisfied by $G_{5,X}=F_5=0$ (Note that the theory with $F_5=-\frac{G_{5,X}}{3 X}\neq 0$, which also sets $b_3=0$, has no\footref{foot ce} tensor and vector modes once we impose Eqn. (\ref{eqn degCondition})). Thirdly, it can be easily checked that $\mathcal{L}_{H_{5A}}$ is a vector-scalar Galileon term. Namely, it is of higher order in the Lagrangian but it has second order equations of motion. Again, no ghosts.

The essential aspect in the  theory (ii) is that it has just the necessary amount of freedom, such that in a subclass within it, the GWs decay to DE may be suppressed: Indeed, in \cite{Creminelli:2018xsv} it was shown that --- in the  case when $c_g(t)\neq 1$ --- the following expression should be negligible (See Eqn. (87) in \cite{Creminelli:2018xsv}), because if the GWs had considerably decayed to DE, we  would have not observed them in first place:
\begin{equation}
    \resizebox{.93\hsize}{!}{ $F_4\Big(4 G_4+X(2 G_{4,X}+3 G_{5,\pi}) \Big)+X F_{4,X}\Big(2 G_4 +X G_{5,\pi} \Big)$}\nonumber
\end{equation}
\vspace{-0.5cm}
\begin{equation}
    \resizebox{1\hsize}{!}{ $+4 G_{4,X}^2+4 G_4 G_{4,XX}+G_{5,\pi}\Big(4 G_{4,X}+2X G_{4,XX}+G_{5,\pi}\Big)=0$}\label{eqn decayConstraint}
\end{equation}
 where we have already used $G_{5,X}=F_5=0$ from the definition of (ii). Note that this constraint is independent of $H$ and $\ddot{\pi}$, thus also independent of the matter content. 
 
As all scalar potentials --- in particular  $F_4(\pi,X)$ --- remain free (while also satisfying Luminality), there are theories in (ii) for which (\ref{eqn decayConstraint}) is satisfied and the GWs decay to DE is suppressed. That is, since Eqn (\ref{eqn decayConstraint}) is  linear in $F_4$, it has solution \cite{Babichev:2024kfo}
\begin{equation}
    F_4=\frac{1}{2X^2}\Big(2 G_4-X(4 G_{4,X}+G_{5,\pi})+\frac{4 J_4(\pi)}{2 G_4+XG_{5,\pi}}\Big),\,\label{ref solF4}
\end{equation}
 where $J_4(\pi)$ is an integration "constant" (with respect to $X$). Notice two essential points to this conclusion: first, the DE--Photon couplings (\ref{eqn BH4A}). They keep free the $F_4(\pi,X)$ function, while also keeping GWs luminal\footnote{Note this critical difference to Beyond Horndeski without the DE--Photon couplings $\mathcal{L}_{\text{BH}_{4A}},\, \mathcal{L}_{\text{H}_{5A}}$: in that case $F_4$ is not free to suppress the decay, because $F_4=\frac{-2 G_{4,X}}{X}$ and $G_5=0$ are already fixed to preserve luminal GWs \cite{Bettoni:2016mij,Ezquiaga:2017ekz,Sakstein:2017xjx,Baker:2017hug,Creminelli:2017sry, Langlois:2017dyl}.}. Thus we can solve $F_4$ as (\ref{ref solF4}). Secondly, the fact that the {\it precise}\footnote{Let us note that in \cite{Babichev:2024kfo} a similar looking Lagrangian to $\mathcal{L}_{\text{H}_{5A}}$ was considered with the aim to suppress the GWs decay while keeping their luminality, $\mathcal{L}_{SVT}^{(3)}\propto \bar{g}_{\alpha\beta}(g,\,\pi,\nabla\pi) \tilde{F}^{\mu\alpha}\tilde{F}^{\nu\beta}\pi_{\mu\nu}$, with $\tilde{F}$ the dual of $F$. However, $\mathcal{L}_{SVT}^{(3)}$ and $\mathcal{L}_{\text{H}_{5A}}$ are fundamentally different. Their quadratic Lagrangians and thus, their vector speeds are related in a matter dependent way, through combinations of $H,\, \ddot{\pi}$. Thus, in accordance with \cite{Babichev:2024kfo} it is not possible to find a matter independent solution to (\ref{eqn decayConstraint}) and (\ref{eqn unitratio})  with $\mathcal{L}_{SVT}^{(3)}$.} $G_5\, F^2 \nabla^2 \pi$ couplings in $\mathcal{L}_{\text{H}_{5A}}$ cannot be removed by a conformal/ disformal transformation of the metric. This is significant: the disformal invariance of the decay, which was proven in \cite{Creminelli:2018xsv} --- and used in the argument in \cite{Babichev:2024kfo} to rule out some BH theories with Scalar-Photon couplings --- does not apply to this case. 
 
 Similarly, to the best of our knowledge, the GWs decay constraint \cite{Creminelli:2018xsv} has only been computed for the BH theory. In particular, additional checks would be needed to rule out the full theory (i) shown above, which includes quadratic BH {\it only} as a particular case.\\
 
 {\normalsize {\bf Conclusions:}} We have shown that 5 sets of self-consistent Dark Energy--Photon couplings are enough to render luminal the GWs in all DHOST theories (with a graviton) that first, are up to cubic in $\nabla^2 \pi$, and second, whose degeneracy conditions are compatible with the sole condition $b_3(\pi,X)=0$. 

 For the cosmologist this means: The Scalar-Tensor theories with $b_3=0$ --- such as the Beyond Horndeski theory (ii) Eqn. (\ref{eqn lagrangianBH4H5}) or (i) --- may be potentially used with minor consideration of the graviton speed, because DE—Photon couplings exist that can take care of the luminality of GWs and the experimental bound (\ref{eqn speedtest}). Naturally,  experimental constraints would be necessary on the DE—Photon couplings proposed in this letter  $\mathcal{L}_{\text{BH}_{4A}},\, \mathcal{L}_{H_{5A}}$ Eqn. (\ref{eqn lagrangianBH4H5}). Indeed, laboratory and astrophysical constraints have been already put on at least the disformal set of DE--Photon couplings  \cite{Brax:2014vva,Brax:2015hma,Babichev:2024kfo}.

 We showed at least one theory --- a subclass of Luminal Beyond Horndeski --- in which the decay of GWs to DE is suppressed on a cosmological background. This aligns with observational evidence and is relevant, because such background is a good description in the bulk of the trajectory of GWs to Earth. We stressed that the essential type of DE--Photon coupling which allows the decay to be suppressed cannot be removed by a conformal/ disformal transformation, and thus, the disformal invariance of the decay --- which was proven in \cite{Creminelli:2018xsv} --- does not apply to this case.

 We also showed some cases of BH and DHOST\footref{foot ce} that remain ruled out by the bound (\ref{eqn speedtest}), as they have no consistent DE--Photon coupling.\\

{\normalsize {\bf Discussion. Recovering GR:}} aside from the experimental constraints on the DE—Photon couplings \cite{Brax:2014vva,Brax:2015hma}, let us note that the Vainshtein screening of the extra scalar mode $\pi$ --- which is essential to align with the precise solar system tests --- remains largely unaffected for at least some of the theories in this letter. 

More precisely, the Vainshtein effect arises when the second order derivative self-interactions become large compared to the kinetic linear term (See \cite{Joyce:2014kja} for a detailed discussion). 

Because the high {\it order of derivatives per field} is essential, let us take the parameter $\alpha=\frac{\partial^2 \pi}{\Lambda^3}$, which may be large with respect to the scalar, photon, metric perturbations and their first derivatives \cite{Koyama:2013paa,Kobayashi:2014ida, Kobayashi:2019hrl} (where $\Lambda$ is the theory's energy scale). The effective action to analyze the Vainshtein effect is thus built keeping $\alpha$ at all orders. This is equivalent to define an order-of-perturbation operator $[\cdot]$ that gives a weight both to fields and derivatives, such that $[\alpha]=0$. In other words, $\alpha^n$ adds zero weight in this perturbative expansion and thus, it will not be truncated. $[\alpha]=0$ amounts to define  $[\pi]=2,\,[\partial]=-1$ and similarly $[A]=[g]=2$, where $\pi,\,A,\, g$ here denote perturbations of their respective fields. Thus the terms that modify the usual {\it quadratic action} $(\partial\, g)^2,\, (\partial\,  A) ^2, \, (\partial \pi)^2$ are in principle $  g\, \alpha^n,\,  (\partial\,  g)^2\, \alpha^n, (\partial\,  A)^2 \, \alpha^n, \pi \, \alpha^n, (\partial\,  \pi)^2 \, \alpha^n $, in accordance with \cite{Koyama:2013paa,Kobayashi:2014ida, Kobayashi:2019hrl}. Notice that all of these quantities are of order 2 under our definition of $[\cdot]$. 

Now, applied to the quadratic DHOST case --- where the DE-Photon couplings are of the type $l_1^{(2)}=(F \partial \pi)^2$ in Eqn. (\ref{eqn l12}) ---  there are no order 2 terms that mix the scalar and the Photon perturbations, because the lowest order mixing --- in the presence of a vanishingly small background vector field --- is $[(\partial  A \, \partial \pi) ^2]=4$, and so the Vainshtein radius remains the same to leading order, independent of the DE-Photon couplings.  Consequently, existing results on the Vainshtein mechanism remain applicable in this case. This includes known constraints and potential issues in {\it e.g.} quadratic DHOST and Beyond Horndeski theories  \cite{Koyama:2013paa,Joyce:2014kja,Kobayashi:2014ida, Kobayashi:2019hrl}. Nevertheless, further studies may help to constrain DHOST theories with DE-Photon couplings. Of special interest are regions of strong magnetic fields, such as Magnetars, where the background vector field becomes relevant or even dominant for these effects. \\

Authors are thankful to S. Ramazanov for valuable discussions. The work on this project
has been supported by Russian Science Foundation grant № 24-72-10110,

\href{https://rscf.ru/project/24-72-10110/}{  https://rscf.ru/project/24-72-10110/}.

\bibliographystyle{ieeetr}

\bibliography{main}

\begin{thebibliography}{10}

\bibitem{LIGOScientific:2017vwq}
B.~P. Abbott {\em et~al.}, ``{GW170817: Observation of Gravitational Waves from
  a Binary Neutron Star Inspiral},'' {\em Phys. Rev. Lett.}, vol.~119, no.~16,
  p.~161101, 2017.

\bibitem{Goldstein:2017mmi}
A.~Goldstein {\em et~al.}, ``{An Ordinary Short Gamma-Ray Burst with
  Extraordinary Implications: Fermi-GBM Detection of GRB 170817A},'' {\em
  Astrophys. J. Lett.}, vol.~848, no.~2, p.~L14, 2017.

\bibitem{Savchenko:2017ffs}
V.~Savchenko {\em et~al.}, ``{INTEGRAL Detection of the First Prompt Gamma-Ray
  Signal Coincident with the Gravitational-wave Event GW170817},'' {\em
  Astrophys. J. Lett.}, vol.~848, no.~2, p.~L15, 2017.

\bibitem{Brax:2014vva}
P.~Brax and C.~Burrage, ``{Constraining Disformally Coupled Scalar Fields},''
  {\em Phys. Rev. D}, vol.~90, no.~10, p.~104009, 2014.

\bibitem{Mironov:2024idn}
S.~Mironov, A.~Shtennikova, and M.~Valencia-Villegas, ``{Reviving Horndeski
  after GW170817 by Kaluza-Klein compactifications},'' {\em Phys. Lett. B},
  vol.~858, p.~139058, 2024.

\bibitem{Bettoni:2016mij}
D.~Bettoni, J.~M. Ezquiaga, K.~Hinterbichler, and M.~Zumalac\'arregui, ``{Speed
  of Gravitational Waves and the Fate of Scalar-Tensor Gravity},'' {\em Phys.
  Rev. D}, vol.~95, no.~8, p.~084029, 2017.

\bibitem{Ezquiaga:2017ekz}
J.~M. Ezquiaga and M.~Zumalac\'arregui, ``{Dark Energy After GW170817: Dead
  Ends and the Road Ahead},'' {\em Phys. Rev. Lett.}, vol.~119, no.~25,
  p.~251304, 2017.

\bibitem{Sakstein:2017xjx}
J.~Sakstein and B.~Jain, ``{Implications of the Neutron Star Merger GW170817
  for Cosmological Scalar-Tensor Theories},'' {\em Phys. Rev. Lett.}, vol.~119,
  no.~25, p.~251303, 2017.

\bibitem{Baker:2017hug}
T.~Baker, E.~Bellini, P.~G. Ferreira, M.~Lagos, J.~Noller, and I.~Sawicki,
  ``{Strong constraints on cosmological gravity from GW170817 and GRB
  170817A},'' {\em Phys. Rev. Lett.}, vol.~119, no.~25, p.~251301, 2017.

\bibitem{Creminelli:2017sry}
P.~Creminelli and F.~Vernizzi, ``{Dark Energy after GW170817 and GRB170817A},''
  {\em Phys. Rev. Lett.}, vol.~119, no.~25, p.~251302, 2017.

\bibitem{Langlois:2017dyl}
D.~Langlois, R.~Saito, D.~Yamauchi, and K.~Noui, ``{Scalar-tensor theories and
  modified gravity in the wake of GW170817},'' {\em Phys. Rev. D}, vol.~97,
  no.~6, p.~061501, 2018.

\bibitem{DESI:2024mwx}
A.~G. Adame {\em et~al.}, ``{DESI 2024 VI: Cosmological Constraints from the
  Measurements of Baryon Acoustic Oscillations},'' 4 2024.

\bibitem{Ye:2024ywg}
G.~Ye, M.~Martinelli, B.~Hu, and A.~Silvestri, ``{Non-minimally coupled gravity
  as a physically viable fit to DESI 2024 BAO},'' 7 2024.

\bibitem{Wolf:2024eph}
W.~J. Wolf, C.~Garc\'\i{}a-Garc\'\i{}a, D.~J. Bartlett, and P.~G. Ferreira,
  ``{Scant evidence for thawing quintessence},'' {\em Phys. Rev. D}, vol.~110,
  no.~8, p.~083528, 2024.

\bibitem{Colgain:2024mtg}
E.~O. Colg\'ain and M.~M. Sheikh-Jabbari, ``{DESI and SNe: Dynamical Dark
  Energy, $\Omega_m$ Tension or Systematics?},'' 12 2024.

\bibitem{Colgain:2024xqj}
E.~O. Colg\'ain, M.~G. Dainotti, S.~Capozziello, S.~Pourojaghi, M.~M.
  Sheikh-Jabbari, and D.~Stojkovic, ``{Does DESI 2024 Confirm $\Lambda$CDM?},''
  4 2024.

\bibitem{Shlivko:2024llw}
D.~Shlivko and P.~J. Steinhardt, ``{Assessing observational constraints on dark
  energy},'' {\em Phys. Lett. B}, vol.~855, p.~138826, 2024.

\bibitem{Cortes:2024lgw}
M.~Cort\^es and A.~R. Liddle, ``{Interpreting DESI's evidence for evolving dark
  energy},'' {\em JCAP}, vol.~12, p.~007, 2024.

\bibitem{Patel:2024odo}
V.~Patel, A.~Chakraborty, and L.~Amendola, ``{The prior dependence of the DESI
  results},'' 7 2024.

\bibitem{Wolf:2024stt}
W.~J. Wolf, P.~G. Ferreira, and C.~Garc\'\i{}a-Garc\'\i{}a, ``{Matching current
  observational constraints with nonminimally coupled dark energy},'' 9 2024.

\bibitem{Langlois:2015cwa}
D.~Langlois and K.~Noui, ``{Degenerate higher derivative theories beyond
  Horndeski: evading the Ostrogradski instability},'' {\em JCAP}, vol.~02,
  p.~034, 2016.

\bibitem{BenAchour:2016fzp}
J.~Ben~Achour, M.~Crisostomi, K.~Koyama, D.~Langlois, K.~Noui, and G.~Tasinato,
  ``{Degenerate higher order scalar-tensor theories beyond Horndeski up to
  cubic order},'' {\em JHEP}, vol.~12, p.~100, 2016.

\bibitem{Langlois:2017mxy}
D.~Langlois, M.~Mancarella, K.~Noui, and F.~Vernizzi, ``{Effective Description
  of Higher-Order Scalar-Tensor Theories},'' {\em JCAP}, vol.~05, p.~033, 2017.

\bibitem{Langlois:2018dxi}
D.~Langlois, ``{Dark energy and modified gravity in degenerate higher-order
  scalar\textendash{}tensor (DHOST) theories: A review},'' {\em Int. J. Mod.
  Phys. D}, vol.~28, no.~05, p.~1942006, 2019.

\bibitem{Mironov:2024pjt}
S.~Mironov and V.~Volkova, ``{Non-singular cosmological scenarios in
  scalar-tensor theories and their stability: a review},'' 9 2024.

\bibitem{Creminelli:2018xsv}
P.~Creminelli, M.~Lewandowski, G.~Tambalo, and F.~Vernizzi, ``{Gravitational
  Wave Decay into Dark Energy},'' {\em JCAP}, vol.~12, p.~025, 2018.

\bibitem{horndeski1974second}
G.~W. Horndeski, ``Second-order scalar-tensor field equations in a
  four-dimensional space,'' {\em International Journal of Theoretical Physics},
  vol.~10, no.~6, pp.~363--384, 1974.

\bibitem{Deffayet:2011gz}
C.~Deffayet, X.~Gao, D.~A. Steer, and G.~Zahariade, ``{From k-essence to
  generalised Galileons},'' {\em Phys. Rev. D}, vol.~84, p.~064039, 2011.

\bibitem{Gleyzes:2014dya}
J.~Gleyzes, D.~Langlois, F.~Piazza, and F.~Vernizzi, ``{Healthy theories beyond
  Horndeski},'' {\em Phys. Rev. Lett.}, vol.~114, no.~21, p.~211101, 2015.

\bibitem{Gleyzes:2014qga}
J.~Gleyzes, D.~Langlois, F.~Piazza, and F.~Vernizzi, ``{Exploring gravitational
  theories beyond Horndeski},'' {\em JCAP}, vol.~02, p.~018, 2015.

\bibitem{Kaluza:1921tu}
T.~Kaluza, ``{Zum Unit\"atsproblem der Physik},'' {\em Sitzungsber. Preuss.
  Akad. Wiss. Berlin (Math. Phys. )}, vol.~1921, pp.~966--972, 1921.

\bibitem{Mironov:2024yqa}
S.~Mironov, M.~Sharov, and V.~Volkova, ``{Linear stability of a time-dependent,
  spherically symmetric background in beyond Horndeski theory and the speed of
  gravity waves},'' 8 2024.

\bibitem{MironovNew}
S.~Mironov, M.~Sharov, and V.~Volkova, ``{Time-dependent, spherically symmetric
  background in Kaluza-Klein compactified Horndeski theory and the speed of
  gravity waves},'' 8 2024.

\bibitem{Mironov:2024umy}
S.~Mironov, A.~Shtennikova, and M.~Valencia-Villegas, ``{Higher derivative
  scalar-vector-tensor theories from Kaluza-Klein reductions of Horndeski
  theory},'' {\em Phys. Rev. D}, vol.~111, no.~2, p.~024028, 2025.

\bibitem{Babichev:2024kfo}
E.~Babichev, C.~Charmousis, B.~Muntz, A.~Padilla, and I.~D. Saltas,
  ``{Horndeski speed tests with scalar-photon couplings},'' 7 2024.

\bibitem{Brax:2015hma}
P.~Brax, C.~Burrage, and C.~Englert, ``{Disformal dark energy at colliders},''
  {\em Phys. Rev. D}, vol.~92, no.~4, p.~044036, 2015.

\bibitem{Joyce:2014kja}
A.~Joyce, B.~Jain, J.~Khoury, and M.~Trodden, ``{Beyond the Cosmological
  Standard Model},'' {\em Phys. Rept.}, vol.~568, pp.~1--98, 2015.

\bibitem{Koyama:2013paa}
K.~Koyama, G.~Niz, and G.~Tasinato, ``{Effective theory for the Vainshtein
  mechanism from the Horndeski action},'' {\em Phys. Rev. D}, vol.~88,
  p.~021502, 2013.

\bibitem{Kobayashi:2014ida}
T.~Kobayashi, Y.~Watanabe, and D.~Yamauchi, ``{Breaking of Vainshtein screening
  in scalar-tensor theories beyond Horndeski},'' {\em Phys. Rev. D}, vol.~91,
  no.~6, p.~064013, 2015.

\bibitem{Kobayashi:2019hrl}
T.~Kobayashi, ``{Horndeski theory and beyond: a review},'' {\em Rept. Prog.
  Phys.}, vol.~82, no.~8, p.~086901, 2019.

\end{thebibliography}

\end{document}